\documentclass[aps,prd,onecolumn,nofootinbib,groupedaddress,superscriptaddress]{revtex4}  
\usepackage{graphicx}
\usepackage{epstopdf}
\usepackage{amsmath}
\usepackage{amsfonts}
\usepackage{amssymb}
\usepackage{appendix}
\usepackage{comment}
\usepackage{bbold}
\usepackage{color}
\usepackage{slashed}
\usepackage{subfigure}
\usepackage{setspace}
\usepackage{footnote}
\usepackage{multirow}
\usepackage{longtable}
\usepackage{braket}
\usepackage{url}

\newcommand{\beq}{\begin{equation}}
\newcommand{\eeq}{\end{equation}}

\begin{document}

\singlespacing

{\hfill NUHEP-TH/21-17, FERMILAB-PUB-21-560-T, IPPP/21/46}

\title{$pp$ Solar Neutrinos at DARWIN}

\author{Andr\'{e} de Gouv\^{e}a} 
\affiliation{Northwestern University, Department of Physics \& Astronomy, 2145 Sheridan Road, Evanston, IL 60208, USA}
\author{Emma McGinness}
\affiliation{University of California Berkeley, Department of Physics, 366 Physics North, Berkeley, CA 94720, USA}
\affiliation{University of California Berkeley, Department of Astronomy, 501 Campbell Hall, Berkeley, CA 94720, USA}
\author{Ivan Martinez-Soler}
\affiliation{Northwestern University, Department of Physics \& Astronomy, 2145 Sheridan Road, Evanston, IL 60208, USA}
\affiliation{Colegio de Física Fundamental e Interdisciplinaria de las Américas (COFI)\\254 Norzagaray street, San Juan, Puerto Rico 00901}
\affiliation{Theoretical Physics Department, Fermilab, P.O. Box 500, Batavia, IL 60510, USA}
\affiliation{Department of Physics \& Laboratory for Particle Physics and Cosmology, Harvard University, Cambridge, MA 02138, USA}
\author{Yuber F. Perez-Gonzalez}
\affiliation{Northwestern University, Department of Physics \& Astronomy, 2145 Sheridan Road, Evanston, IL 60208, USA}
\affiliation{Colegio de Física Fundamental e Interdisciplinaria de las Américas (COFI)\\254 Norzagaray street, San Juan, Puerto Rico 00901}
\affiliation{Theoretical Physics Department, Fermilab, P.O. Box 500, Batavia, IL 60510, USA}
\affiliation{Institute for Particle Physics Phenomenology, Durham University, South Road, Durham, United Kingdom.}

\begin{abstract}

The DARWIN collaboration recently argued that DARWIN (DARk matter WImp search with liquid xenoN) can collect, via neutrino--electron scattering,  a large, useful sample of solar $pp$-neutrinos, and measure their survival probability with sub-percent precision. We explore the physics potential of such a sample in more detail. We estimate that, with 300~ton-years of data, DARWIN can also measure, with the help of current solar neutrino data, the value of $\sin^2\theta_{13}$, with the potential to exclude $\sin^2\theta_{13}=0$ close to the three-sigma level.  We explore in some detail how well DARWIN can constrain the existence of a new neutrino mass-eigenstate $\nu_4$ that is quasi-mass-degenerate with $\nu_1$ and find that DARWIN's sensitivity supersedes that of all current and near-future searches for new, very light neutrinos. In particular, DARWIN can test the hypothesis that $\nu_1$ is a pseudo-Dirac fermion as long as the induced mass-squared difference is larger than $10^{-13}$~eV$^2$, one order of magnitude more sensitive than existing constraints. Throughout, we allowed for the hypotheses that DARWIN is filled with natural xenon or $^{136}$Xe-depleted xenon. 

\end{abstract}

\maketitle

\section{Introduction}
\label{sec:intro}

Multi-ton-scale, next-generation dark matter experiments are expected to collect significant statistics of atmospheric and solar neutrinos. The DARWIN collaboration recently argued that DARWIN (DARk matter WImp search with liquid xenoN) can collect a large, useful sample of solar $pp$-neutrinos, measured via elastic neutrino--electron scattering  \cite{DARWIN:2020bnc}. There, they argued that the survivial probability of $pp$-neutrinos can be measured with sub-percent precision and that one can measure the Weinberg angle at low momentum transfers with 10\% precision, independent from the values of the neutrino oscillation parameters. Here, we explore other neutrino-physics-related information one can obtain from a high-statistics, high-precision measurement of the $pp$-neutrino flux. 

In a nutshell, $pp$-neutrinos are produced in the solar core via proton-proton fusion: $p+p\to{\rm ^2H}+e^++\nu$. The vast majority of neutrinos produced by the fusion cycle that powers our Sun are produced via proton-proton fusion. $pp$-neutrinos have the lowest energy among all solar neutrino ``types'' (other types include $pep$-neutrinos, $^7{\rm Be}$-neutrinos, $^8{\rm B}$-neutrinos, and CNO-neutrinos) and are characterized by a continuous spectrum that peaks around 300~keV and terminates around 420~keV. Theoretically, the $pp$-neutrino flux is known at better than the percent level \cite{Bahcall:2000nu} given they are created early in the $pp$-fusion cycle -- they are the first link in the chain -- and their flux is highly correlated with the photon flux, measured with exquisite precision. For the sake of comparison, the flux of $^7{\rm Be}$-neutrinos and $^8{\rm B}$-neutrinos, which provide virtually all information on the particle-physics properties of solar neutrinos, can be computed at, approximately, the 6\% and 12\% level, respectively \cite{Serenelli:2011py,Vinyoles:2016djt}.  The $pp$-neutrino flux has been directly measured, independent from the other flux-types, by the Borexino collaboration \cite{BOREXINO:2014pcl}, with 10\% precision. 

A percent-level measurement of the $pp$-neutrino flux is expected to be sensitive to new-physics effects in neutrino physics that are also at the percent level. This includes, for example, effects from the so-called reactor angle $\theta_{13}$ -- not new physics but very small for solar neutrinos -- and the presence of new neutrino states or neutrino interactions. Furthermore, the fact that $pp$-neutrinos have energies that are significantly lower than those of the other solar neutrino types renders them especially well-suited to constrain (or discover) new, very long oscillation lengths associate to very small new neutrino mass-squared differences. These searches are expected to add significantly to our ability to test the hypothesis that the neutrinos are pseudo-Dirac fermions \cite{Wolfenstein:1981kw,Petcov:1982ya,Bilenky:1983wt} (for relevant recent discussions, see, for example, \cite{deGouvea:2009fp,Donini:2011jh,DeGouvea:2020ang,Martinez-Soler:2021unz}). Here, as far as new-physics hypotheses are concerned,  we concentrate on the search for new, very light neutrino states.

In Sec.~\ref{sec:DARWIN}, we review the relevant features of the proposed DARWIN experiment and provide information on how we simulate and analyze DARWIN data on $pp$-neutrinos. In Sec.~\ref{sec:3nus}, we show that a percent-level measurement of the $pp$-neutrino flux allows for a ``solar-neutrinos-only'' measurement of $\sin^2\theta_{13}$. In Sec.~\ref{sec:4nus}, we compute the sensitivity of DARWIN to the hypothesis that there is a fourth neutrino with a mass $m_4$ that is quasi-degenerate with the mass of the first neutrino state, $m_1$ (in the Appendix, we discuss how this can be generalized). We concentrate on the region of parameter space where the new mass-squared difference is $10^{-13}~{\rm eV^2}\lesssim |m_4^2-m_1^2| \lesssim 10^{-6}$~eV$^2$. We add some concluding remarks in Sec.~\ref{sec:conclusion}.

\section{DARWIN as a Low-Energy Solar Neutrino Experiment}
\label{sec:DARWIN}
\setcounter{equation}{0}

DARWIN is projected to be a large -- 40~tons fiducial volume --  liquid xenon time-projection chamber, aimed at searching for weakly interacting massive particles (WIMP) in the GeV to TeV mass range \cite{DARWIN:2016hyl} via elastic WIMP--nucleon scattering. It will inevitably be exposed to a large flux of solar and atmospheric neutrinos and is large enough that solar-neutrino scattering events will occur at an observable rate.

According to \cite{DARWIN:2020bnc}, DARWIN is expected to collect a sample of almost ten thousand $pp$-neutrinos per year via elastic neutrino--electron scattering:
\begin{equation}
\nu_{\alpha}+e^-\to\nu_{\alpha}+e^-,
\end{equation}
where $\alpha=e,\mu,\tau$ is the  flavor of the incoming neutrino. The flavor of the outgoing neutrinos is, of course, never observed. For $pp$-neutrino energies, the cross section for $\nu_ee$-scattering is around six times larger than that of $\nu_ae$-scattering, $a=\tau,\mu$ and the differences between the cross sections for $\nu_{\mu}e$-scattering and $\nu_{\tau}e$-scattering are negligible. At leading order in the weak interactions, the differential cross section in the rest frame of the electron is 
\begin{equation}
\frac{{\rm d}\sigma}{{\rm d}T} (\nu_{\alpha}+e^-\to\nu_{\alpha}+e^-)=\frac{2G_F^2m_e}{\pi}\left[a_{\alpha}^2+b_{\alpha}^2\left(1-\frac{T}{E_{\nu}}\right)^2-a_{\alpha}b_{\alpha}\frac{T}{E_{\nu}}\right],
\end{equation}
where $T$ is the kinetic energy of the recoil electron, $E_{\nu}$ is the incoming neutrino energy, $m_e$ is the electron mass and $G_F$ is the Fermi constant. The dimensionless couplings $a_{\alpha},b_{\alpha}$ are
\begin{equation}
a_e = -\frac{1}{2}-\sin^2\theta_W, ~~~b_e= -\sin^2\theta_W;~~~ a_a=\frac{1}{2}-\sin^2\theta_W, ~~~ b_a= -\sin^2\theta_W,
\end{equation}
where $\theta_W$ is the weak mixing angle. DARWIN measures the kinetic energy spectrum of the recoil electrons. 

If filled with natural xenon, one expects a large number of electron-events in the energy range of interest from the double-beta decays of $^{136}\rm Xe$. These events are a powerful source of background for solar-neutrino studies and, according to \cite{DARWIN:2020bnc}, obviate the study of solar neutrinos with energies higher than 1~MeV. They are a powerful nuisance for measurements of the $^7\rm Be$-neutrino flux and have a significant but not decisive impact on the measurement of the $pp$-neutrinos (around a 30\% decrease in the precision with which the overall $pp$-neutrino flux can be measured \cite{DARWIN:2020bnc}). The reason one can measure the $pp$-neutrino flux in spite of the $^{136}\rm Xe$ background is that the shape of this particular background is well known and the experiment can detect events over a large range of recoil-electron energies, effectively measuring it with excellent precision. There is the possibility of filling DARWIN with liquid xenon depleted of the double-beta-decaying $^{136}\rm Xe$ isotope. This would allow the study of higher energy solar neutrinos. Here we consider these two different scenarios, i.e., the $^{136}\rm Xe$-depleted version of DARWIN and the one where the abundance of $^{136}\rm Xe$ agrees with natural expectations. 

Other than the background from $^{136}\rm Xe$, for $pp$-neutrinos, the double electron capture decay of $^{124}\rm Xe$ leads to two narrow peaks at 37~keV and 10~keV \cite{DARWIN:2020bnc} and, at higher recoil energies, radioactive backgrounds from the detector components and the liquid volume supersede the $pp$-neutrino events for recoil kinetic energies above 200~keV or so. When simulating DARWIN data, we restrict our sample to events with recoil kinetic energies below 220~keV and assume that, in this energy range, the only sources of background are those from $^{136}\rm Xe$ and $^{124}\rm Xe$. We simulate the backgrounds using the results published in \cite{DARWIN:2020bnc}. When analyzing the simulated data, we marginalize over the normalization of the two $^{124}\rm Xe$ lines, which we treat as free parameters, and the normalization of the $^{136}\rm Xe$ recoil spectrum, which we assume is independently measured with 0.1\% precision. We assume the shape of the $^{136}\rm Xe$ recoil spectrum is known with infinite precision. For the $^{136}\rm Xe$-depleted version of DARWIN, we assume the $^{136}\rm Xe$-background is $1\%$ of the background presented in~\cite{DARWIN:2020bnc}. We organize the simulated data into recoil-kinetic-energy bins with 10~keV width, consistent with the recoil-kinetic-energy resolution quoted in \cite{DARWIN:2020bnc}, starting at 1~keV. We use a simple $\chi^2$-test in order to address questions associated to the sensitivity of DARWIN to different parameters and in order to combine simulated DARWIN data with those from other experiments.

\section{Testing The Three-Massive-Neutrinos Paradigm}
\label{sec:3nus}
\setcounter{equation}{0}

In the absence of more new physics, existing data reveals that the neutrino weak-interaction-eigenstates $\nu_{\alpha}$, $\alpha=e,\mu,\tau$, are linear combinations of the neutrino mass-eigenstates $\nu_i$ with mass $m_i$, $i=1,2,3$:
\begin{equation}
\nu_\alpha = U_{\alpha i}\nu_i,
\end{equation}
where the $U_{\alpha i}$, $\alpha=e,\mu,\tau$, $i=1,2,3$, define the elements of a unitary matrix. Here, we are only interested in solar neutrinos so all accessible observables are sensitive to $|U_{ei}|^2$, $i=1,2,3$. These, in turn, are parameterized with two mixing angles, $\theta_{12}$ and $\theta_{13}$. Following the parameterization of the Particle Data Group \cite{Zyla:2020zbs},
\begin{equation}
|U_{e2}|^2=\sin^2\theta_{12}\cos^2\theta_{13},~~~~~|U_{e3}|^2=\sin^2\theta_{13},
\label{eq:Ue2Ue3}
\end{equation}
and unitarity uniquely determines the third matrix-element-squared: $|U_{e1}|^2=1-|U_{e2}|^2-|U_{e3}|^2$. Combined fits to the existing data reveal that the two independent mass-squared differences are $\Delta m^2_{21}\equiv m_2^2-m_1^2\sim 10^{-4}$~eV$^2$ and $|\Delta m^2_{31}|\equiv m_3^2-m_1^2\sim 10^{-3}$~eV$^2$. For more precise values see, for example, \cite{Esteban:2020cvm}.\footnote{See also \url{http://www.nu-fit.org}.} While $\Delta m^2_{21}$ is defined to be positive, the sign of $\Delta m^2_{31}$ is still unknown; for our purposes here, it turns out, this is irrelevant. The two mixing parameters of interest have been measured quite precisely. According to \cite{Esteban:2020cvm}, at the one-sigma level,
\begin{equation}
\sin^2\theta_{12}=0.304^{+0.013}_{-0.012},~~~~\sin^2\theta_{13}=0.02221^{+0.00068}_{-0.00062}.
\label{eq:nufit}
\end{equation}
The experiments that contribute most to these two measurements are qualitatively different. $\theta_{12}$ is best constrained by solar neutrino experiments -- and is often referred to as the ``solar angle'' -- while $\theta_{13}$ is best constrained by reactor antineutrino experiments -- and is often referred to as the ``reactor angle.''

We are interested in the solar $pp$-neutrinos. These have a continuous energy spectrum that peaks around 300~keV and terminates at around 420~keV. The matter-potential $V=\sqrt{2}G_FN_e$, where $G_F$ is the Fermi constant and $N_e$ is the electron number-density, inside the Sun is $V_{\odot}<2\times 10^{-5}$~(eV$^2$/MeV) so, for neutrino energies $E<0.420$~MeV, $|\Delta m^2_{21}|/(2E),|\Delta m^2_{31}|/(2E)\gg V_{\odot}$.\footnote{For example, at the center of the Sun, for neutrino energies less than 420~keV, the ``matter equivalent'' of $\sin^22\theta_{12}$ differs from its vacuum counterpart by less than one percent.} This, in turn, implies that, given what is known about the neutrino mass-squared differences, matter effects can be neglected. Including the fact that, for all practical purposes, solar neutrinos lose flavor coherence as they find their way from the Sun to the Earth, it is trivial to show that the $\nu_e$ survival probability is energy independent and given by
\begin{equation}
P_{ee}=|U_{e1}|^4+|U_{e2}|^4+|U_{e3}|^4.
\label{eq:Pee3f}
\end{equation}
On the other hand, solar neutrino experiments cannot distinguish $\nu_{\mu}$  from $\nu_{\tau}$ -- the neutrino energies are too small -- but are potentially sensitive to the combination $P_{ea}\equiv P_{e\mu}+P_{e\tau}$. In the three-massive-neutrinos paradigm
\begin{equation}
P_{ea} = 1- P_{ee}.
\end{equation}
Given our current knowledge of mixing parameters, for $pp$-neutrinos, we can indirectly infer that $P_{ee}=0.552\pm0.025$, naively combining the uncertainties in Eq.~(\ref{eq:nufit}) in quadrature. 

\begin{figure}[ht]
\includegraphics[width=0.95\textwidth]{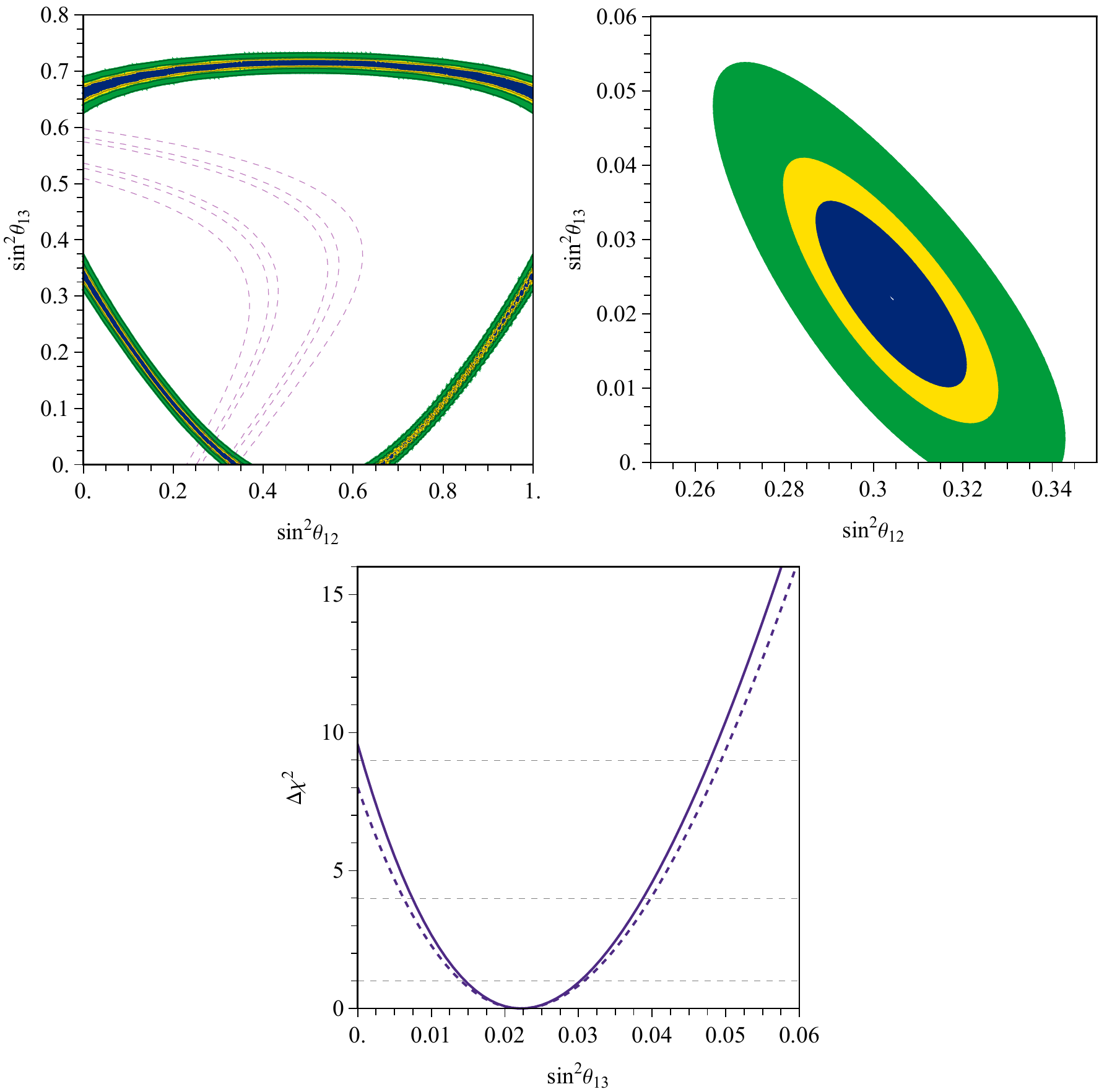}
\caption{Top: One-, two- and three-sigma allowed regions of the $\sin^2\theta_{12}\times \sin^2\theta_{13}$ parameter space assuming DARWIN can measure $P_{ee}=0.552$ at the one percent level, excluding (left) and including (right) external constraints on the neutrino-mixing parameters from other solar experiments. The tiny empty ellipse in the right-hand panel indicates the best-fit point. The open regions bound by dashed lines (left-hand panel) represent  one-, two- and three-sigma results from current $^8$B neutrino experiments, as discussed in the text. Bottom: Solar-only $\chi^2$ as a function of $\sin^2\theta_{13}$, marginalized over $\sin^2\theta_{12}$, assuming 300-ton-years of simulated DARWIN data. The full line corresponds to the assumption of a depleted background, while the dashed line is obtained including the expected natural $^{136}{\rm Xe}$ background.} 
\label{fig:3nus}
\end{figure}

According to \cite{DARWIN:2020bnc}, after 20 ton-years of exposure, DARWIN can measure $P_{ee}$ with better than 1\% accuracy. Assuming the three-massive-neutrinos paradigm, this can be converted into a measurement of the relevant mixing parameters. Fig.~\ref{fig:3nus}(top,left) depicts the allowed region of the $\sin^2\theta_{12}\times \sin^2\theta_{13}$ parameter space assuming DARWIN can measure $P_{ee}$ for $pp$-neutrinos at the 1\% level, and assuming the best-fit value is $P_{ee}=0.552$. There is very strong degeneracy between different values of $\sin^2\theta_{12}$ and $\sin^2\theta_{13}$, for obvious reasons. The degeneracies can be lifted by including constraints from other neutrino experiments. 

It is interesting to investigate how well one can constrain neutrino-mixing parameters using only solar-neutrino data. In order to estimate that, we add to the hypothetical $pp$-neutrino measurement from DARWIN current information from $^8$B neutrinos, mostly from the Super-Kamiokande and SNO experiments, see \cite{SNO:2011hxd,Super-Kamiokande:2016yck} and references therein. These provide the strongest constraints on $\sin^2\theta_{12}$. Here, we address this in a simplified but accurate way \cite{Nunokawa:2006ms}, postulating that $^8$B experiments measure 
\begin{equation} 
(P_{ee})_{^8\rm B,~average} = (1-|U_{e3}|^2) \left[0.9|U_{e2}|^2+ 0.1|U_{e1}|^2\right] +|U_{e3}|^4,
\end{equation}
with 4\% accuracy, consistent with the current uncertainty on $\sin^2\theta_{12}$, mostly constrained by high-energy solar neutrino data. $(1-|U_{e3}|^2)\times 0.9$ (or $(1-|U_{e3}|^2)\times 0.1$) is the average probability that a $^8$B neutrino arrives at the surface of the Earth as a $\nu_2$ (or a $\nu_1$). When the $^8$B data is treated as outlined above, it translates into the open regions bound by dashed lines in Fig.~\ref{fig:3nus}(top,left). Strong matter effects lead to the boomerang-shaped allowed region of the parameter space and restrict the parameter space to values of $\sin^2\theta_{12}\lesssim 0.5$. The results of the joint $pp-^8$B analysis are depicted in Fig.~\ref{fig:3nus}(top,right). All degeneracies present in the $pp$-neutrino data are lifted and one is constrained to small values of $\sin^2\theta_{13}$ and $\sin^2\theta_{12}<0.5$. 
 
The combined $^8$B and DARWIN data can rule out $\sin^2\theta_{13}=0$ with some precision. This is important; it implies that a hypothetical DARWIN measurement of the $pp$-neutrino flux, combined with the current $^8$B solar neutrino data, can measure $\sin^2\theta_{13}$ in a way that is independent from all non-solar measurements. The marginalized $\chi^2$ as a function of $\sin^2\theta_{13}$ is depicted in  Fig.~\ref{fig:3nus}(bottom), for 300~ton-years of simulated DARWIN data and the current $^8$B sollar neutrino data. On average, if the $pp$-neutrino flux can be measured at the percent level, we expect to measure $\sin^2\theta_{13}$ at the 35\% level and rule out $\sin^2\theta_{13}=0$ at almost the three-sigma level. Here we consider the two scenarios outlined earlier, one with natural xenon (dashed line), the other with $^{136}$Xe-depleted xenon (solid line).
 
 The precision on $\sin^2\theta_{13}$ obtained above is not comparable to that of the current measurement of $\sin^2\theta_{13}$, Eq.~(\ref{eq:nufit}). However, these measurements are qualitatively different. The most precise measurements of $\sin^2\theta_{13}$ come from reactor antineutrino experiments and a baseline of order 1~km \cite{RENO:2018dro,DayaBay:2018yms,DoubleChooz:2019qbj}. The estimate discussed above is a ``solar only'' measurement, i.e., it exclusively makes use of measurements of neutrinos (and not antineutrinos) produced in the Sun. Current measurements of $\sin^2\theta_{13}$ that make use of neutrinos (as opposed to antineutrinos), from T2K and NOvA, are much less precise (at the 50\%, see \cite{NOvA:2016kwd,T2K:2017rgv}). Looking further into the future, the DUNE experiment, for example, is expected to independently measure the ``neutrino-only'' value of $\sin^2\theta_{13}$ at the 20\% level \cite{deGouvea:2017yvn} (or worse, depending on the assumptions made in the analysis).

\section{Beyond the Three-Massive-Neutrinos Paradigm}
\label{sec:4nus}
\setcounter{equation}{0}

The fact that the $pp$-neutrino flux can be computed with great precision, combined with the sub-MeV $pp$-neutrino energies, allows a high-statistics measurement of the $pp$-neutrino flux to meaningfully search for phenomena beyond the thee-massive-neutrinos paradigm. Here we concentrate on testing the hypothesis that the neutrinos produced in the Sun have a nonzero probability of behaving as ``sterile neutrinos'' $\nu_s$, characterized by their lack of participation in charged-current and neutral-current weak interactions. 

We first discuss, in Sec.~\ref{sec:4nusMI},  the case where the oscillation probabilities are energy-independent for the energies of interest, as in the case of the thee-massive-neutrinos paradigm discussed in Sec.~\ref{sec:3nus}. In particular, we test the hypothesis that $P_{ee}+P_{ea}=1$ for $pp$-neutrinos.  Then, in Sec.~\ref{sec:4nus4}, we compute DARWIN's ability to constrain the hypothesis that there is a fourth neutrino $\nu_4$ and that its mass is quasi-degenerate with $m_1$.  

\subsection{Model-Independent Considerations}
\label{sec:4nusMI}

As discussed in Sec.~\ref{sec:DARWIN}, we are interested in the shape and normalization of the electron recoil-energy spectrum from neutrino--electron elastic scattering. The differential cross-section for $\nu_e$ and $\nu_a$  scattering are different, both in normalization and shape and hence, in principle, one can obtain independent information on both $P_{ee}$ and $P_{ea}$. 

\begin{figure}[ht]
\includegraphics[width=0.95\textwidth]{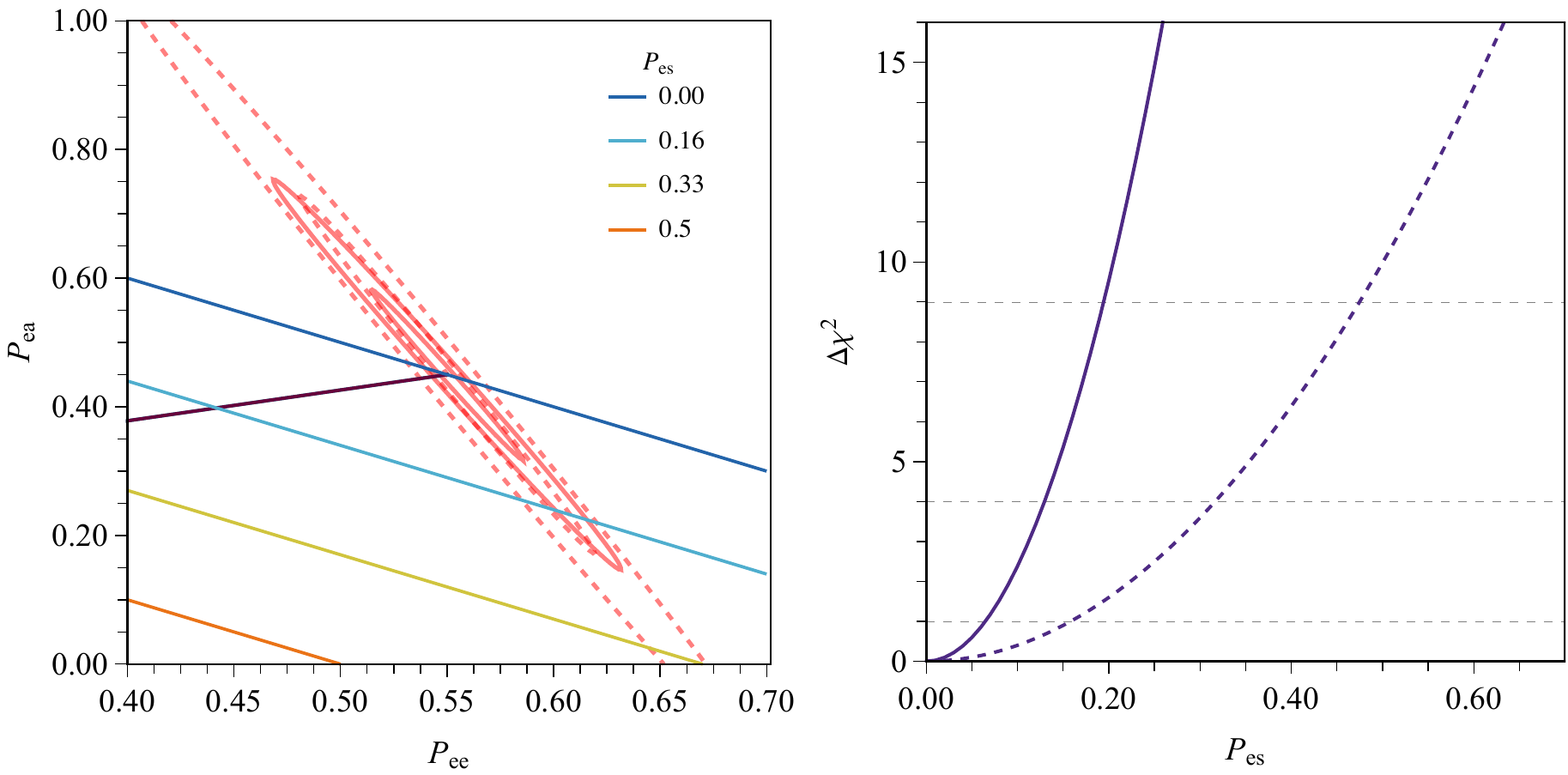}
\caption{Left: One- and three-sigma allowed region of the $P_{ee}\times P_{ea}$-plane, for 300 ton-years of simulated DARWIN data. The diagonal lines correspond to constant $P_{es}\equiv 1-P_{ee}-P_{ea}$ values. The burgundy  line segment with positive slope corresponds to the values of $(P_{ee},P_{ea})$ accessible via Eqs.~(\ref{eq:Pave}). Right: Marginalized $\chi^2$ as a function of $P_{es}$. The full line correspond to the assumption of a depleted background, while the dashed line is obtained considering no cuts in the $^{136}{\rm Xe}$ background.} 
\label{fig:PeePea}
\end{figure}

We simulate and analyze 300 ton-years of DARWIN $pp$-data, as discussed in Sec.~\ref{sec:DARWIN}, and attempt to measure $P_{ee}$ and $P_{ea}$ independently. The results are depicted in Fig.~\ref{fig:PeePea}(left) for both the natural xenon (dashed) and the $^{136}$Xe-depleted (solid) hypotheses. Strong departures from $P_{ee}+P_{ea}=1$ are allowed and the ``natural'' data are not capable of ruling out $P_{ea}=0$ at the three-sigma confidence level. The ``depleted'' data can rule out $P_{ea}=0$ at the five-sigma confidence level. For both scenarios, one can constrain the departure of $P_{ee}+P_{ea}$ from one, which we interpret as the oscillation probability into sterile neutrinos $P_{es}\equiv 1-P_{ee}-P_{ea}$. The colorful diagonal lines in  Fig.~\ref{fig:PeePea}(left) correspond to different constant values of $P_{es}$. Fig.~\ref{fig:PeePea}(right) depicts $\chi^2$ as a function of $P_{es}$, marginalized over $P_{ee}$ and restricing $P_{es}$ to non-negative values for both scenarios. If DARWIN data are consistent with the three-active-neutrinos paradigm, they will be capable of constraining $P_{es}<0.35$ at the two-sigma confidence level even if DARWIN is filled with natural xenon.

\subsection{Fourth-Neutrino Hypothesis}
\label{sec:4nus4}

We explore in more detail the scenario where there is one extra neutrino mass-eigenstate $\nu_4$ with mass $m_4$. In this case, the interaction eigenstates are, including $\nu_s$, related to the four mass-eigenstates via a $4\times 4$ unitary matrix $U_{\alpha i}$, $\alpha=e,\mu,\tau,s$, $i=1,2,3,4$. We will concentrate on the scenario where, among the four $U_{si},$ only $U_{s1}$ and $U_{s4}$ are potentially nonzero.\footnote{It is easy to generalize this analysis assuming that only one of the $U_{sj}$, $j=1,2,3$, and $U_{s4}$ are potentially nonzero. We spell this out in the Appendix.} In this case, we can parameterize the $|U_{e i}|^2$ entries of the mixing matrix using three mixing angles $\theta_{12},\theta_{13},\theta_{14}$. Eqs.~(\ref{eq:Ue2Ue3}) are still valid, along with
\begin{equation}
|U_{e1}|^2=\cos^2\theta_{12}\cos^2\theta_{13}\cos^2\theta_{14},~~~~~|U_{e4}|^2=\cos^2\theta_{12}\cos^2\theta_{13}\sin^2\theta_{14}.
\label{eq:Ue1Ue4}
\end{equation}
It is easy to check that $\sum_{i=1}^4|U_{ei}|^2=1$. The non-zero ``sterile'' entries of the mixing matrix are
\begin{equation}
|U_{s1}|^2=\sin^2\theta_{14},~~~~~|U_{s4}|^2=\cos^2\theta_{14}.
\label{eq:Ue1Ue4s}
\end{equation}
Given the quasi-two-flavors nature of these solar neutrino oscillations, to be discussed momentarily, the entire physical parameter space is spanned by either fixing $\Delta m^2_{41}>0$ and allowing $\sin^2\theta_{14}\in[0,1]$ or allowing both signs for $\Delta m^2_{41}$ and restricting $\sin^2\theta_{14}\in[0,0.5]$ in such a way that $\nu_4$ is always ``mostly sterile.'' Here, the former convention -- to fix the sign of $\Delta m^2_{41}>0$ -- is most convenient. With this choice, when $\sin^2\theta_{14}\in[0,0.5]$,  the heaviest of the two quasi-degenerate states (i.e., $\nu_4$)  is mostly sterile, when $\sin^2\theta_{14}\in[0.5,1]$, the lightest among the two quasi-degenerate states (i.e., $\nu_1$) is mostly sterile. For historical reasons, we will refer to $\sin^2\theta_{14}\in[0,0.5]$ as the light side of the parameter space and $\sin^2\theta_{14}\in[0.5,1]$ as the dark side \cite{deGouvea:2000pqg}. 

We are interested in the hypothesis that $\Delta m^2_{41}\ll \Delta m^2_{21}$ and outside the reach of all current neutrino experiments. In this case, the current neutrino oscillation data constrain the oscillation parameters $\Delta m^2_{21}$, $\Delta m^2_{31}$, $\sin^2\theta_{12}$, and $\sin^2\theta_{13}$ exactly as in the three-massive-neutrinos paradigm. Furthermore, builiding on the discussion in Sec.~\ref{sec:3nus}, it is easy to conclude that the oscillation probabilities of interest $P_{e\alpha}$, $\alpha=e,a,s$, are only functions of $\sin^2\theta_{12},\sin^2\theta_{13},\sin^2\theta_{14}$, and $\Delta m^2_{41}$. Further taking advantage of the fact that $|\Delta m^2_{21}|/(2E),|\Delta m^2_{31}|/(2E)\gg V_{\odot}$, it is straightforward to compute
\begin{eqnarray}
P_{ee}&=&|U_{e2}|^4+|U_{e3}|^4+(1-|U_{e2}|^2-|U_{e3}|^2)^2P_{ee}^{2f}(\Delta m^2_{41},\sin^2\theta_{14},V^{\rm eff}), \label{eq:Pee4} \\
P_{es}&=&(1-|U_{e2}|^2-|U_{e3}|^2)(1-P_{ee}^{2f}(\Delta m^2_{41},\sin^2\theta_{14},V^{\rm eff})), \\
P_{ea}&=&1-P_{ee}-P_{es},
\end{eqnarray}
where $P_{ee}^{2f}$ is the survival probability obtained in the scenario where there are only two flavors, $\nu_e^{2f}$  and $\nu_s^{2f}$, characterized by the mass-squared difference $\Delta m^2_{41}$ and the mixing angle $\theta_{14}$, defined via $\nu_e^{2f}=\cos\theta_{14}\nu_1+\sin\theta_{14}\nu_4$. Inside $P_{ee}^{2f}$, the matter potential is replaced by an effective matter potential $V^{\rm eff}$. It takes into account the neutral-current contribution to the matter potential $V_{NC}=-\sqrt{2}/2G_FN_n$, where $N_n$ is the neutron number density in the medium, while the charged-current contribution is rescaled by $(1-|U_{e2}^2-|U_{e3}|^2)=\cos^2\theta_{13}\cos^2\theta_{12}$:
\begin{equation}
V^{\rm eff} = \sqrt{2} G_F \left(N_e\cos^2\theta_{13}\cos^2\theta_{12}-\frac{1}{2}N_n\right). \label{eq:Veff}
\end{equation} 
In the sun, the position-dependency of the electron and neutron number densities are slightly different  \cite{Bahcall:2000nu}.\footnote{There are relatively more neutrons in the center of the sun relative to its edges. This is due to the fact that most of the solar helium is concentrated in the core.} In the sun's core, $N_n$ is around 50\% of $N_e$ and $V^{\rm eff}$ is slightly less than one half of the standard matter potential in the Sun.

Eq.~(\ref{eq:Pee4}) allows us to estimate, in very general terms, the impact of the sterile neutrinos. For $P^{2f}_{ee}=1$, we recover the three-active-neutrinos result, $P_{ee}=\cos^4\theta_{12}\cos^4\theta_{13}+\sin^4\theta_{12}\cos^4\theta_{13}+\sin^4\theta_{13}$, Eq.~(\ref{eq:Pee3f}). On the other hand, for $P^{2f}_{ee}=0$, $P_{ee}=\sin^4\theta_{12}\cos^4\theta_{13}+\sin^4\theta_{13}$ such that, given the current knowledge of oscillation parameters, 
\begin{equation}
P_{ee}\in [0.09,0.55].
\end{equation}
$P_{es}$ values, on the other hand, are allowed to be as small as zero and as large as 0.68.

In Sec.~\ref{sec:4nusMI}, we discussed that, very generically, DARWIN can rule out $P_{es}<0.35$ at the two-sigma level. The situation here is more constrained as $P_{ee}, P_{ea}, P_{es}$ are not only required to add up to one but depend on the same oscillation parameters. We proceed to discuss the sensitivity of DARWIN to the new oscillation parameters $\Delta m^2_{41},\sin^2\theta_{14}$ by taking advantage of the fact that the properties of $P_{ee}^{2f}$ are well known (see, for example, \cite{Giunti:1053706}). 

\begin{figure}[ht]
\includegraphics[width=0.45\textwidth]{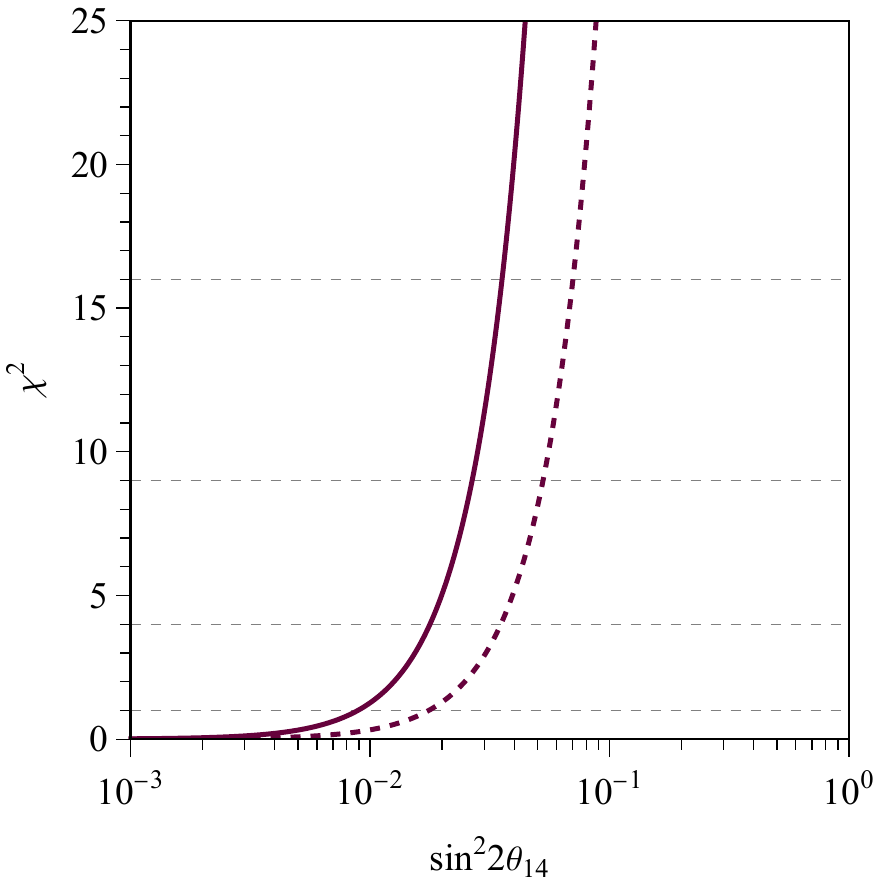}
\caption{$\chi^2$ as a function of $\sin^2\theta_{14}$ for 300 ton-years of simulated DARWIN data, in the regime when matter effects are not significant and the new oscillatory effects driven by $\Delta m^2_{41}$ average out. The full line corresponds to the assumption of a depleted background, while the dashed line is obtained considering no cuts in the $^{136}{\rm Xe}$ background.} 
\label{fig:chitheta14}
\end{figure}

For large-enough values of $\Delta m^2_{41}$, $P_{ee}^{2f}$ is well approximated by averaged-out vacuum oscillations: 
\begin{equation}
P_{ee}^{2f,\rm ave} = 1-\frac{1}{2}\sin^22\theta_{14}.
\label{eq:P_ee2fave}
\end{equation}  
This occurs, keeping in mind we are interested in energies below 420~keV, for $\Delta m^2_{41}\gtrsim 10^{-5}$~eV$^2$, when the solar matter effects can be ignored. In this case, 
\begin{eqnarray}
P_{ee}&=&0.55-0.23\sin^22\theta_{14}, \nonumber \\
P_{es}&=&0.34\sin^22\theta_{14}, \label{eq:Pave} \\
P_{ea}&=&0.45-0.11\sin^22\theta_{14}. \nonumber
\end{eqnarray}
Here, it is impossible to distinguish the light from the dark side of the parameter space since the oscillation probabilities are invariant under $\sin^2\theta_{14}\leftrightarrow 1-\sin^2\theta_{14}$. Varying $\sin^22\theta_{14}\in[0,1]$, Eqs.~(\ref{eq:Pave}) define a line segment in the $P_{ee}\times P_{ea}$-plane, depicted in Fig.~\ref{fig:PeePea}(left) -- burgundy line with positive slope -- keeping in mind the segment extends to $P_{ee}$ values below 0.4. Fig.~\ref{fig:chitheta14} depicts $\chi^2$ as a function of $\sin^2\theta_{14}$ in the regime where Eqs.~(\ref{eq:Pave}) are a good approximation, for 300 ton-years of simulated DARWIN data, for both the natural (dashed) and $^{136}$Xe-depleted (solid) scenarios. In this analysis, and in the upcoming analyses discussed this subsection, we assume that $\sin^2\theta_{13}$ and $\sin^2\theta_{12}$ are known with infinite precision. This is, currently, a good approximation for $\sin^2\theta_{13}$ and will be a good approximation for $\sin^2\theta_{12}$ once data from the JUNO experiment is analyzed \cite{JUNO:2015zny}. Similar results were recently presented and discussed in \cite{Goldhagen:2021kxe}. Where are assumptions agree, the estimated sensitivity also agrees. 

For intermediate values of $\Delta m^2_{41}$, $P_{ee}^{2f}$ is well describe by the strong MSW effect in the adiabatic regime. In this case, for a range of energies, 
\begin{equation}
P_{ee}^{2f,\rm adiabatic} = \sin^2\theta_{14}.
\label{eq:P_ee2fadiabatic}
\end{equation}  
For $pp$-neutrinos, this occurs for, very roughly, $\sin^2\theta_{14}\gtrsim 10^{-3}$ and $10^{-9}\lesssim \Delta m^2_{41}/(\rm eV^2)\lesssim 10^{-6}$. Under these conditions,
\begin{eqnarray}
P_{ee}&=&0.09+0.46\sin^2\theta_{14}, \nonumber \\
P_{es}&=&0.68-0.68\sin^2\theta_{14}, \\
P_{ea}&=& 0.27+0.22\sin^2\theta_{14}. \nonumber
\end{eqnarray}
Here, oscillation probabilities are very different in the light and dark sides. In particular, in the light side of the parameter space $P_{ee}$ ($P_{es}$) is small (large) and increases (decreases) linearly with $\sin^2\theta_{14}$. If DARWIN data are consistent with three-active neutrinos, in this region of parameter space, small values of $\sin^2\theta_{14}$ will be excluded while large values of $\sin^2\theta_{14}$ are allowed. 

For small-enough values of $\Delta m^2_{41}$, $P_{ee}^{2f}$ is well described by the strong MSW effect in the very non-adiabatic regime and turns out to be well approximated by vacuum oscillations, 
\begin{equation}
P_{ee}^{2f,\rm ave} = 1-\sin^22\theta_{14}\sin^2\left(\frac{\Delta m^2_{41}L}{4E}\right).
\label{eq:P_ee2fvac}
\end{equation}  
This occurs, for $pp$-neutrinos, for $\Delta m^2_{41}\lesssim 10^{-9}$~eV$^2$. In this case, 
\begin{eqnarray}
P_{ee}&=&0.55-0.46\sin^22\theta_{14}\sin^2\left(\frac{\Delta m^2_{41}L}{4E}\right), \nonumber \\
P_{es}&=&0.68\sin^22\theta_{14}\sin^2\left(\frac{\Delta m^2_{41}L}{4E}\right), \\
P_{ea}&=&0.45-0.22\sin^22\theta_{14}\sin^2\left(\frac{\Delta m^2_{41}L}{4E}\right). \nonumber
\end{eqnarray}
Here, again, it is impossible to distinguish the light from the dark side of the parameter space. Given the average Earth--Sun distance $L=1.5\times 10^{12}$~m, the oscillation phase is 
\begin{equation}
\frac{\Delta m^2_{41}L}{4E} = 4.8\left(\frac{\Delta m^2_{41}}{10^{-12}~\rm eV^2}\right)\left(\frac{400~\rm keV}{E}\right),
\end{equation}
so we expect the vacuum oscillations to average out for $\Delta m^2_{41}\gtrsim 10^{-11}$~eV$^2$. This means that, for $10^{-11}\lesssim \Delta m^2_{41}/(\rm eV^2)\lesssim 10^{-9}$, the oscillation probabilities are well described by Eqs.~(\ref{eq:Pave}).

\begin{figure}[ht]
\includegraphics[width=0.85\textwidth]{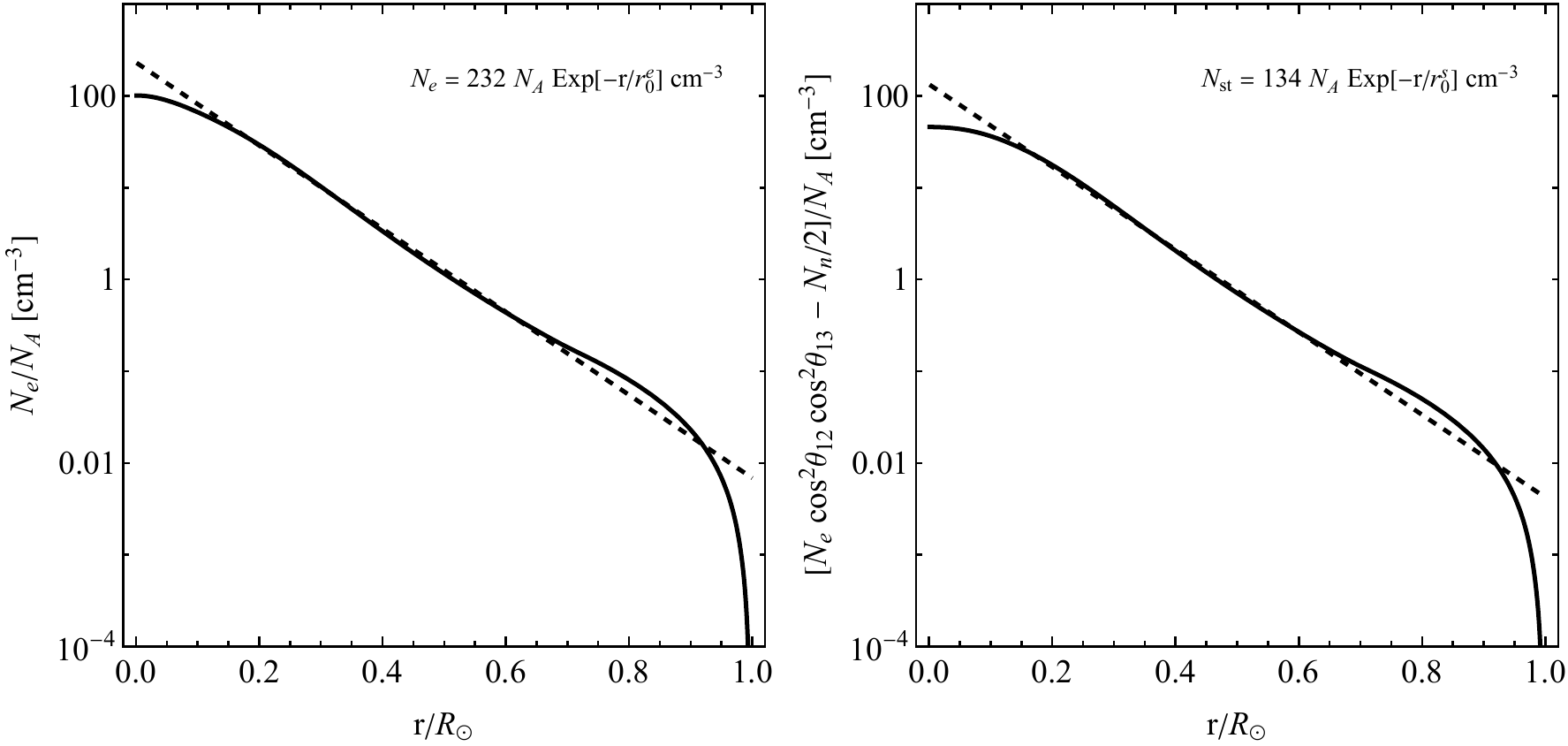}
\caption{Solar matter potential for active (left) and sterile (right) neutrinos -- the scenario of interest here -- as function of the distance from the center in units of the Solar radius, from the B16-GS98 Solar Model~\cite{Vinyoles:2016djt}. We also present our fitted exponential forms, where $r_0^e=R_{\odot}/10.43$, and $r_0^s=R_{\odot}/10.37$, in dashed lines.} 
\label{fig:InterpMPs}
\end{figure}

\begin{figure}[ht]
\includegraphics[width=0.55\textwidth]{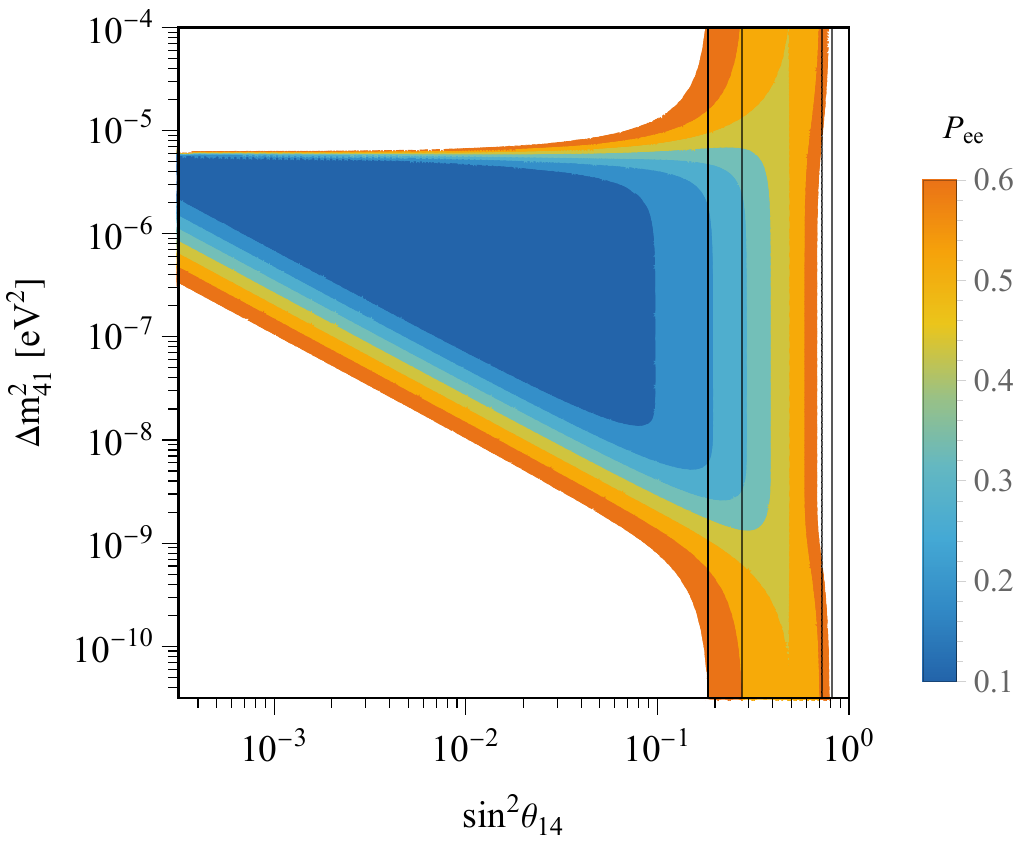}
\caption{Contours of constant $P_{ee}$ in the $\Delta m^2_{41}\times \sin^2\theta_{14}$-plane for $E=300$~keV. $\sin^2\theta_{12},\sin^2\theta_{13}$ are fixed to their best-fit values, Eq.~(\ref{eq:nufit}). The vertical lines correspond to constant values of the averaged-out vacuum oscillation probability.}
\label{fig:Pee300}
\end{figure}

Fig.~\ref{fig:Pee300} depicts contours of constant $P_{ee}$ in the $\Delta m^2_{41}\times \sin^2\theta_{14}$-plane for $E_{\nu}=300$~keV. The other parameters are fixed to their current best-fit values, Eq.~(\ref{eq:nufit}). We assume the matter potential is spherically symmetric and drops exponentially, $V^{\rm eff}\propto e^{-r/r_0^s}$. We fit information from the prediction of the B16-GS98 solar model~\cite{Vinyoles:2016djt} and obtain $r_0^s=R_{\odot}/10.37$ where $R_{\odot}=6.96\times 10^{11}$~m is the average radius of the Sun; see Fig.~\ref{fig:InterpMPs} for a comparison of the matter potential in the standard case (left) and in the scenario of interest here (right, labeled sterile neutrino). 

Under these circumstances, $P_{ee}^{2f}$ can be computed exactly \cite{Petcov:1987zj}. For simplified pedagogical discussions see, for example, \cite{deGouvea:2004gd,Giunti:1053706}.  We assume all solar neutrinos are produced in the exact center of the Sun; we explicitly verified that the results we get are very similar to the results we would have obtained by integrating over the region where $pp$-neutrinos are produced. The region where matter effects are strong and the adiabatic condition holds correspond to the vertical sides of the constant $P_{ee}$ regions that form quasi-triangles. The ``return'' to vacuum oscillations at low and high values of $\Delta m^2_{41}$ is highlighted by the vertical, dark lines, which correspond to constant values of the averaged-out vacuum oscillation probability. For a detailed discussion of the boundary between the adiabatic and non-adiabatic transition, including $L$ dependent effects, see \cite{Friedland:2000rn}.

We simulate 300 ton-years of DARWIN data consistent with the three-massive-neutrinos paradigm and assuming the true values of $\sin^2\theta_{12}$ and $\sin^2\theta_{13}$ are the ones in Eq.~(\ref{eq:nufit}). We restrict our discussion to values of $\Delta m^2_{41}<10^{-6}$~eV$^2$. Larger values are constrained by measurements of higher-energy solar neutrinos; these constraints have been explored in \cite{deHolanda:2003tx,deHolanda:2010am}, along with a detailed discussion of the oscillation probabilities. The expressions we derive here are contained in the analyses of  \cite{deHolanda:2003tx,deHolanda:2010am} if one explores them in the appropriate regime. 

\begin{figure}[ht]
\includegraphics[width=0.475\textwidth]{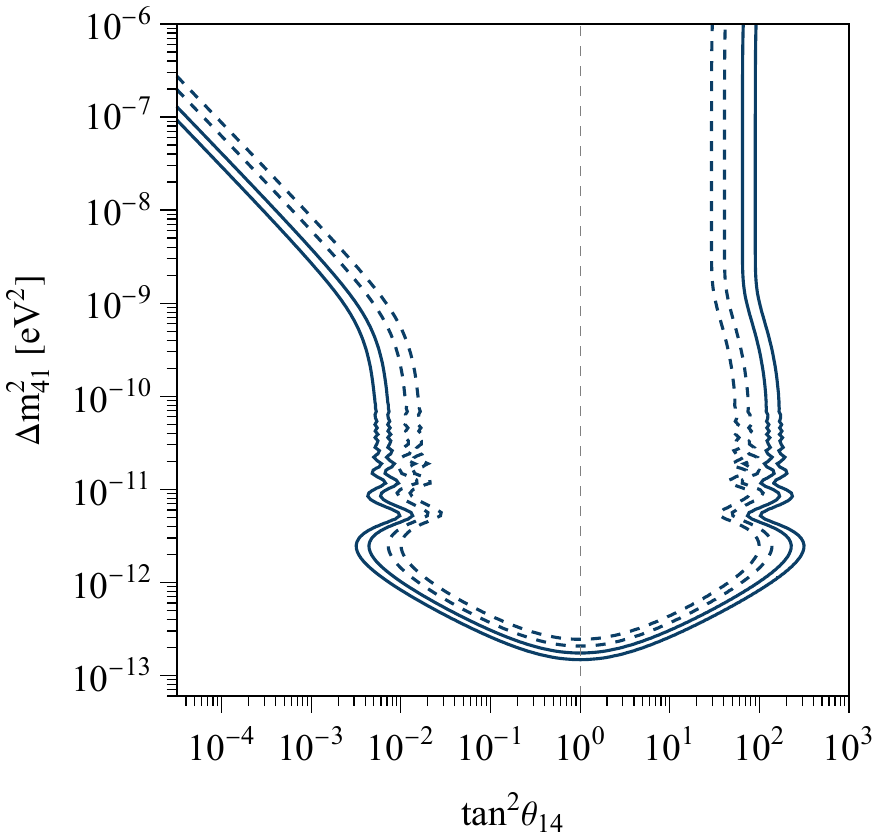}
\caption{Two- and three-sigma sensitivity of 300~ton-years of DARWIN data to a light sterile neutrino in the $\Delta m^2_{41}\times \tan^2\theta_{14}$--plane. The dashed lines correspond to the assumption of a full $^{136}$Xe background, while the full lines are for the depleted case.}
\label{fig:2D}
\end{figure}

Fig.~\ref{fig:2D} depicts the region of the $\tan^2\theta\times \Delta m^2_{41}$--plane inside of which 300 ton-years of DARWIN data is sensitive, at the 90\% confidence level, to the fourth neutrino for both the natural-xenon scenario (dashed line) and the depleted-$^{136}$Xe scenario (solid). On a log-scale, the contour is symmetric relative to $\tan^2\theta_{14}=1$ when one cannot distinguish the light from the dark side of the parameter space \cite{deGouvea:2000pqg}, a feature one readily observes, as advertised, for small values of $\Delta m^2_{41}$. The impact of nontrivial matter effects is also readily observable. For larger values of the $\Delta m^2_{41}$, the sensitivity to small mixing angles is expected to ``shut-off'' quickly -- see Fig.~\ref{fig:Pee300} -- and would return to values similar to those around $\Delta m^2_{41}\sim 10^{-10}$~eV$^2$, minus the tiny wiggles. 

The low energies of the $pp$-neutrinos combined with the long Earth--Sun distance render DARWIN a specially powerful probe of the hypothesis that neutrinos are pseudo-Dirac fermions. This is the hypothesis that there are right-handed neutrinos coupled to the left-handed lepton doublets and the Higgs  doublet via a tiny Yukawa coupling $y$ and that lepton number is only slightly violated. In these scenarios, each of the neutrino mass eigenstates is ``split'' into two quasi-degenerate Majorana fermions, each a 50--50 mixture of an active neutrino (from the lepton doublet) and a sterile neutrino (the right-handed neutrino). The mass splitting is small enough that, for most applications, the two quasi-degenerate state act as one Dirac fermion. Pseudo-Dirac neutrinos reveal themselves via active--sterile oscillations associated with very large mixing and very small mass-squared differences. 

\begin{figure}[ht]
\includegraphics[width=0.45\textwidth]{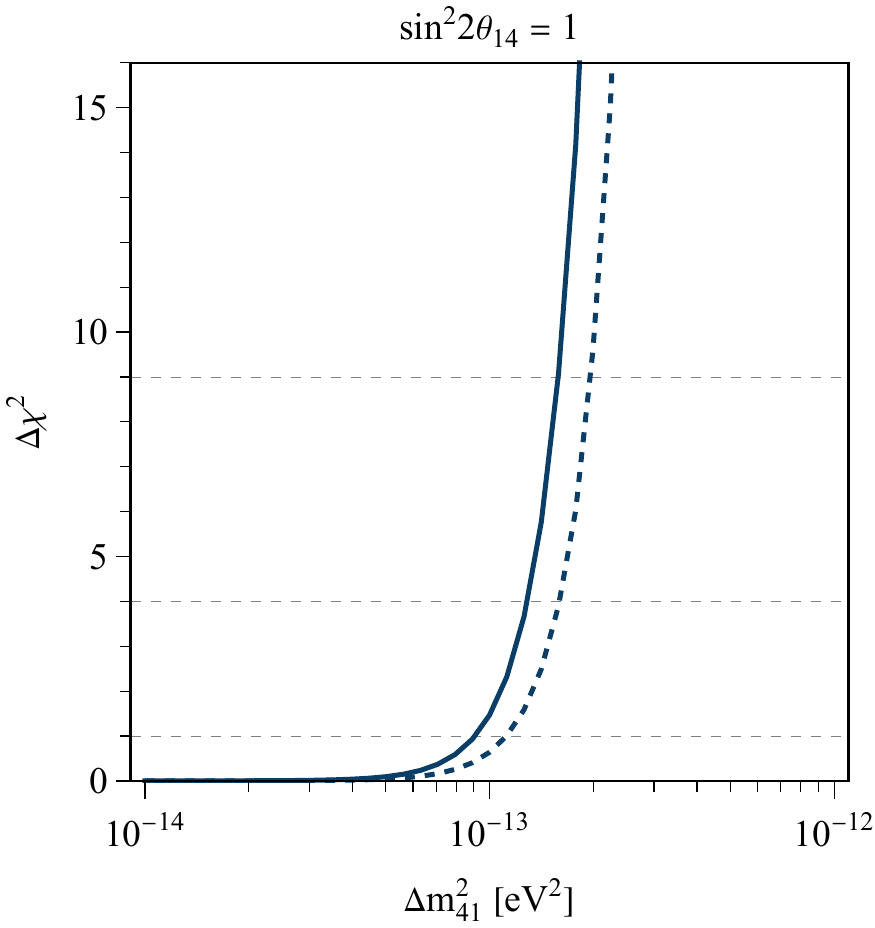}
\caption{Darwin sensitivity to the quadratic mass difference $\Delta m^2_{41}$ in the case of maximal mixing, $\sin^2\theta_{14}=0.5$. The full line correspond to the assumption of a depleted background, while the dashed line is obtained considering no cuts in the $^{136}{\rm Xe}$ background.}
\label{fig:chisq}
\end{figure}

In the language introduced here, a pseudo-Dirac neutrino corresponds to $\sin^22\theta_{14}=1$ (maximal mixing) and the small mass splitting leads to a nonzero $\Delta m^2_{41}=4\epsilon m_1$ where $m_1\pm\epsilon$ are the masses of the two quasi-degenerate states (here $\nu_1$ and $\nu_4$), $m_1$ is the Dirac mass, proportional to the neutrino Yukawa coupling and $\epsilon$ characterizes the strength of the lepton-number violating physics. Fig.~\ref{fig:chisq} depicts $\chi^2$ as a function of $\Delta m^2_{41}$ for $\sin^22\theta_{14}=1$ associated with 300 ton-years of simulated DARWIN data for both the natural  xenon (dashed) and the $^{136}$Xe-depleted (solid) scenarios, assuming the data are consistent with no new neutrino states. Current solar neutrino data exclude $\Delta m^2_{41}$ values larger than $10^{-12}$~eV$^2$  \cite{deGouvea:2009fp,Donini:2011jh} so DARWIN can extend the sensitivity to $\Delta m^2_{41}$ -- and $\epsilon$ -- by an order of magnitude. 

\section{Concluding Remarks}
\label{sec:conclusion}
\setcounter{equation}{0}

Next-generation WIMP-dark-matter-search experiments will be exposed to a large-enough flux of solar neutrinos that neutrino-mediated events are unavoidable. The DARWIN collaboration recently argued that DARWIN can collect a large, useful sample of solar $pp$-neutrinos, detected via elastic neutrino--electron scattering  \cite{DARWIN:2020bnc}, and measure the survival probability of $pp$-neutrinos with sub-percent precision. Here we explored the physics potential of such a sample in more detail, addressing other concrete neutrino-physics questions and exploring whether one can also extract information from a precise measurement of the shape of the differential $pp$-neutrino flux. 

We estimate that, with 300~ton-years of data, DARWIN can not only measure the survival probability of $pp$-neutrinos with sub-percent precision but also determine, with the help of current solar neutrino data, the value of $\sin^2\theta_{13}$, with the potential to exclude $\sin^2\theta_{13}=0$ close to the three sigma level. Such a $pp$-neutrino sample would allow one to perform a ``neutrinos-only'' (and solar-neutrinos-only) measurement of $\sin^2\theta_{13}$ and $\sin^2\theta_{12}$. Such a measurement can be compared with, for example, reactor-based ``antineutrinos-only'' measurements of the same mixing parameters and allow for nontrivial tests of the CPT-theorem and other new physics scenarios. 

DARWIN can also test the hypothesis that $pp$-neutrinos are oscillating into a combination of active and sterile neutrinos. We estimate that DARWIN data can exclude the hypothesis that the $pp$-neutrinos are ``disappearing'' in an energy independent way -- assuming their data are consistent with the three-active-neutrinos paradigm -- especially if the experiment manages to fill their detector with $^{136}$Xe-depleted xenon.   

We explored in some detail how well DARWIN can constrain the existence of a new neutrino mass-eigenstate $\nu_4$ (mass $m_4$) that is quasi-degenerate and mixes with $\nu_1$, i.e, $\Delta m^2_{41}\ll \Delta m^2_{21}$, $U_{s1},U_{s4}\neq0$, $U_{s2}=U_{s3}=0$. Our estimated sensitivity is depicted in Fig.~\ref{fig:2D}. It supersedes that of all current and near-future searches for new, very light neutrinos. In particular, DARWIN can test the hypothesis that $\nu_1$ is a pseudo-Dirac fermion as long as the induced mass-squared difference is larger than $10^{-13}$~eV$^2$. This is one order of magnitude more sensitive than existing constraints \cite{deGouvea:2009fp,Donini:2011jh}.   

Throughout, we allowed for the hypotheses that DARWIN is filled with natural xenon or $^{136}$Xe-depleted xenon. We find that while the sensitivity of the experiment with natural xenon is outstanding, a $^{136}$Xe-depleted setup is significantly more sensitive when it comes to the measurements and searches discussed here. In our discussions, we did not include time-dependent effects (the seasonal and day-night effects). These can impact the sensitivity to new neutrino states within a subset of the parameter space explored here. They would not significantly modify the results discussed here but provide extra handles on the new physics.

\appendix
\section{Other Fourth-Neutrino Scenarios}

We restricted our fourth-neutrino analyses to one new neutrino mass-eigenstate $\nu_4$ and allowed for only ``1--4'' sterile mixing, i.e., $U_{s1}=\sin\theta_{14}$, $U_{s4}=\cos\theta_{14}$ while $U_{s2}=U_{s3}=0$. Here we discuss some simple generalizations. 

The case of one new neutrino mass-eigenstate $\nu_5$ that is quasi-degenerate with $\nu_2$ and only ``2--5'' sterile mixing would also be parameterized by a mass-squared difference $\Delta m^2_{52}$ (positive-definite), assumed to be much smaller than $\Delta m^2_{21}$, $\Delta m^2_{51}$ and $|\Delta m^2_{53}|$, and one mixing angle $\theta_{25}$:
\begin{align}
U_{e1}^2 &= \cos^2\theta_{12}\cos^2\theta_{13}, & U_{e2}^2 &= \sin^2\theta_{12}\cos^2\theta_{13}\cos^2\theta_{25}, & U_{e3}^2 &= \sin^2\theta_{13}, & U_{e5}^2 &= \sin^2\theta_{12}\cos^2\theta_{13}\sin^2\theta_{25}, \\
U_{s1}^2 &= 0, & U_{s2}^2 &= \sin^2\theta_{25},  & U_{s3}^2 &= 0, & U_{s5}^2 &= \cos^2\theta_{25}.
\end{align}
Similar to Eq.~(\ref{eq:Pee4}), and the equivalent expressions for $P_{ea}$ and $P_{es}$, here
\begin{eqnarray}
P_{ee}&=&|U_{e1}|^4+|U_{e3}|^4+(1-|U_{e1}|^2-|U_{e3}|^2)^2P_{ee}^{2f}(\Delta m^2_{52},\sin^2\theta_{25},V_{25}^{\rm eff}), \label{eq:Pee5} \\
P_{es}&=&(1-|U_{e1}|^2-|U_{e3}|^2)(1-P_{ee}^{2f}(\Delta m^2_{52},\sin^2\theta_{25},V_{25}^{\rm eff})), \\
P_{ea}&=&1-P_{ee}-P_{es},
\end{eqnarray}
where $P_{ee}^{2f}$ is the survival probability obtained in the scenario where there are only two flavors, $\nu_e^{2f}$  and $\nu_s^{2f}$, characterized by the mass-squared difference $\Delta m^2_{52}$ and the mixing angle $\theta_{25}$, defined via $\nu_e^{2f}=\cos\theta_{25}\nu_2+\sin\theta_{25}\nu_5$. Here, the effective matter potential is
\begin{equation}
V_{25}^{\rm eff} = \sqrt{2} G_F \left(N_e\cos^2\theta_{13}\sin^2\theta_{12}-\frac{1}{2}N_n\right). \label{eq:V25}
\end{equation} 

On the other hand, the case of one new neutrino mass-eigenstate $\nu_6$ that is quasi-degenerate with $\nu_3$ and only ``3--6'' sterile mixing would also be parameterized by a mass-squared difference $\Delta m^2_{63}$ (positive-definite), assumed to be much smaller than $\Delta m^2_{21}$, $|\Delta m^2_{61}|$ and $|\Delta m^2_{62}|$, and one mixing angle $\theta_{36}$:
\begin{align}
U_{e1}^2 &= \cos^2\theta_{12}\cos^2\theta_{13}, & U_{e2}^2 &= \sin^2\theta_{12}\cos^2\theta_{13}, & U_{e3}^2 &= \sin^2\theta_{13}\cos^2\theta_{36}, & U_{e5}^2 &= \sin^2\theta_{13}\sin^2\theta_{36}, \\
U_{s1}^2 &= 0, & U_{s2}^2 &= 0,  & U_{s3}^2 &= \sin^2\theta_{36}, & U_{s5}^2 &= \cos^2\theta_{36}.
\end{align}
Similar to Eq.~(\ref{eq:Pee4}), and the equivalent expressions for $P_{ea}$ and $P_{es}$, here
\begin{eqnarray}
P_{ee}&=&|U_{e1}|^4+|U_{e2}|^4+(1-|U_{e1}|^2-|U_{e2}|^2)^2P_{ee}^{2f}(\Delta m^2_{63},\sin^2\theta_{36},V_{36}^{\rm eff}), \label{eq:Pee6} \\
P_{es}&=&(1-|U_{e1}|^2-|U_{e2}|^2)(1-P_{ee}^{2f}(\Delta m^2_{63},\sin^2\theta_{36},V_{36}^{\rm eff})), \\
P_{ea}&=&1-P_{ee}-P_{es},
\end{eqnarray}
where $P_{ee}^{2f}$ is the survival probability obtained in the scenario where there are only two flavors, $\nu_e^{2f}$  and $\nu_s^{2f}$, characterized by the mass-squared difference $\Delta m^2_{63}$ and the mixing angle $\theta_{36}$, defined via $\nu_e^{2f}=\cos\theta_{36}\nu_3+\sin\theta_{36}\nu_6$. Here, the effective matter potential is
\begin{equation}
V_{36}^{\rm eff} = \sqrt{2} G_F \left(N_e\sin^2\theta_{13}-\frac{1}{2}N_n\right). \label{eq:V36}
\end{equation} 

Qualitatively, the three scenarios -- 1--4, 2--5, 3--6 -- are identical modulo relabelings of the mixing parameters. Quantitatively, however, there are significant differences. The coefficients of the $P_{ee}^{2f}$ term in Eqs.~(\ref{eq:Pee4}), (\ref{eq:Pee5}), (\ref{eq:Pee6}) are, respectively, $(\cos^2\theta_{12}\cos^2\theta_{13})^2\sim 0.5$, $(\sin^2\theta_{12}\cos^2\theta_{13})^2\sim 0.1$, and $(\sin^2\theta_{13})^2\sim 0.0005$. These numbers define the maximum deviation of $P_{ee}$ from expectations from the three-massive-neutrinos paradigm, $P_{ee}\sim 0.55$. Hence, very generically, 1--4 effects can be very strong, as discussed in the text, 2--5 effects are at most of order 20\%, and 3--6 effects are at the permille level. On the other hand, the effective potentials are also quantitatively very different. The charged-current contribution to $V^{\rm eff}_{25}$ (Eq.~(\ref{eq:V25})) is suppressed relative to the neutral-current one by a factor $\sin^2\theta_{12}\cos^2\theta_{13}\sim 0.3$. Since $N_n/N_e$ varies between, roughly, 0.5 and less than 0.1 between the center of the Sun and its edge, $V^{\rm eff}_{25}$ is significantly smaller than $V_{\odot}$, almost vanishing at the Sun's core, when the charged- and neutral-current contributions, accidentally, almost cancel out one another. $V^{\rm eff}_{36}$ (Eq.~(\ref{eq:V36})), instead, is solidly dominated by the neutral-current matter potential since the charged-current contribution is suppressed by $\sin^2\theta_{13}\sim 0.02$. Not only is it smaller than $V_{\odot}$, it has the opposite sign, a fact that qualitatively impact the behavior of $P_{ee}^{2f}$.

The scenario where all neutrinos are pseudo-Dirac fermions is equivalent to the  combination of the 1--4, 2--5, and 3--6 scenarios spelled out above (see, for example, \cite{deGouvea:2009fp}). Note that such a combination is straight forward; the effects of the different contributions simply ``add up'' without too much interference, as long as the new mass-squared differences are ``isolated enough,'' i.e., the three new mass-squared differences $\Delta m^2_{41}$,   $\Delta m^2_{52}$, and $\Delta m^2_{63}$ are much smaller than all other mass-squared differences. For example, 
\begin{eqnarray}
P_{ee} &=& (\cos^2\theta_{12}\cos^2\theta_{13})^2P_{ee}^{2f}(\Delta m^2_{41},\sin^2\theta_{14},V_{14}^{\rm eff}) +(\sin^2\theta_{12}\sin^2\theta_{13})^2P_{ee}^{2f}(\Delta m^2_{52}\sin^2\theta_{25},V_{25}^{\rm eff}) + \nonumber \\
&+& (\sin^2\theta_{13})^2P_{ee}^{2f}(\Delta m^2_{63},\sin^2\theta_{36},V_{36}^{\rm eff}),
\end{eqnarray}
where $V_{14}^{\rm eff}$ is given by Eq.~(\ref{eq:Veff}).

\section*{Acknowledgements}
We thank Pedro Machado for discussions of potential uses of DARWIN data for neutrino physics. This work was supported in part by the US Department of Energy (DOE) grant \#de-sc0010143 and in part by the NSF grant PHY-1630782. The document was prepared using the resources of the Fermi National Accelerator Laboratory (Fermilab), a DOE, Office of Science, HEP User Facility. Fermilab is managed by Fermi Research Alliance, LLC (FRA), acting under Contract No. DE-AC02-07CH11359.  This material is based upon work supported by the NSF grant AST-1757792, a Research Experience for Undergraduates grant awarded to the Center for Interdisciplinary Exploration and Research in Astrophysics (CIERA) at Northwestern University. IMS is supported by the Faculty of Arts and Sciences of Harvard University.


\bibliographystyle{apsrev-title}
\bibliography{darwin_pp.bib}{}

\end{document}